\documentstyle[12pt,aaspp4,flushrt]{article}

\newcommand{\be}{\begin{equation}}
\newcommand{\ee}{\end{equation}}
\newcommand{\mstar}{M_\star}
\newcommand{\wb}{\widetilde b}
\newcommand{\bfr}{{\bf r}}
\newcommand{\bfv}{{\bf v}}
\newcommand{\bfb}{{\bf b}}
\newcommand{\bfJ}{{\bf J}}
\newcommand{\bfc}{{\bf c}}
\newcommand{\bfM}{{\bf M}}

\begin{document}

\title{\bf Resonant relaxation in protoplanetary disks}
\author{Scott Tremaine}
\affil{Princeton University Observatory\\Peyton Hall, Princeton, NJ 08544}

\medskip

\begin{abstract}
Resonant relaxation is a novel form of two-body relaxation that arises in
nearly Keplerian disks such as protoplanetary disks. Resonant
relaxation does not affect the semimajor axes of the particles, but enhances
relaxation of particle eccentricities and inclinations. The equilibrium state
after resonant relaxation is a Rayleigh distribution, with the mean-square
eccentricity and inclination inversely proportional to mass. The rate of
resonant relaxation depends strongly on the precession rate of the disk. If
the precession due to the disk's self-gravity is small compared to the total
precession, then the relaxation is concentrated near the secular resonance
between each pair of interacting bodies; on the other hand if the precession
rate is dominated by the disk's self-gravity then relaxation occurs through
coupling to the large-scale low-frequency $m=1$ normal modes of the disk.
Depending on the disk properties, resonant relaxation may be either stronger
or weaker than the usual non-resonant relaxation.

\end{abstract}

\section{Introduction}

\label{sec:intro}

\noindent
The formation of planets from a disk of planetesimals is largely determined by
two closely related processes: physical collisions and gravitational
relaxation (e.g. \cite{SW88}, \cite{LS93}). Collisions result from close
two-body encounters and drive the evolution of the mass spectrum of the
planetesimals, while relaxation arises mainly from distant two-body encounters
and drives the evolution of their phase-space distribution. The relative rates
of the two processes are determined by the Safronov number, $\theta\sim
(v_e/v_r)^2$, where $v_e$ is the surface escape speed and $v_r$ is the rms
random velocity of the planetesimals; relaxation dominates when $\theta\gg 1$.
Gravitational relaxation has two main effects on the random velocity of a
particle in a disk: scattering drives a stochastic random walk towards larger
values of $v_r$, while dynamical friction damps $v_r$. In general, scattering
dominates for small particles while dynamical friction dominates for massive
ones, as expected by equipartition.

The usual derivation of the rate of gravitational relaxation treats the
evolution of the particle orbit as a sequence of uncorrelated two-body
encounters with other particles. Such derivations neglect the effect of
the gravitational field of the central star, but should be accurate in the
early stages of planetesimal accumulation, so long as $v_r\gg nr_H$, where $n$
is the mean motion and $r_H=r(m/\mstar)^{1/3}$ is the Hill radius of the
particle; here $m$ and $\mstar$ are the masses of the particle and the
central star, and $r$ is the orbital radius (\cite{I90}).

This papers investigates a qualitatively different type of gravitational
relaxation, resonant relaxation, which we have already investigated in the
context of spherical stellar systems (\cite{RT96}).  Resonant relaxation
arises in addition to the usual (``non-resonant'') relaxation discussed above,
when the potential in which the particles orbit is nearly Keplerian. To
introduce the concept, let us imagine a time-exposure of a particulate
Keplerian disk, in which the exposure time is much longer than the orbital
period $t_{\rm orb}$ but shorter than the characteristic precession time
$t_{\rm prec}$. Each particle appears in the photograph as an eccentric
ellipse or wire. These wires exert torques on one another which remain roughly
constant for $\sim t_{\rm prec}$; however, after several times $t_{\rm prec}$
the configuration of the wires and the resulting torques will be quite
different. These torques cause the angular momentum vectors of the
particles---and thus their eccentricities and inclinations---to random walk,
with the duration of one ``step'' in the walk being roughly $t_{\rm
prec}$. However, the energies or semimajor axes are unaffected because the
potential from the wires is nearly stationary.  A closely related process is
resonant friction, which damps the eccentricities and inclinations of massive
particles without affecting their semimajor axes. The combined effects of
resonant relaxation and resonant friction drive the phase-space distribution
of particles towards the maximum-entropy state consistent with fixed particle
energies and total angular momentum. The main goal of the paper is to estimate
the resonant rates of excitation and damping of protoplanet eccentricities, to
compare these briefly with the non-resonant rates, and to demonstrate that in
some cases resonant relaxation dominates the eccentricity evolution. 

Resonant relaxation is a close cousin of secular perturbation theory in
celestial mechanics, which likewise averages the Hamiltonian over times much
longer than $t_{\rm orb}$. Thus an alternative name for the process we
are examining would be the oxymoron ``secular relaxation''.

Nearly Keplerian disks can be parametrized by two dimensionless numbers
\be
S(r)\equiv -{g(r)\over n(r)}{\mstar\over\pi r^2\Sigma(r)}=-{g(r)n(r)r\over\pi
G\Sigma(r)},  \qquad Q(r)\equiv
1.07{\sigma_r \over n(r)r}{\mstar\over \pi r^2\Sigma(r)};
\label{eq:sdef}
\ee
here $g=\dot\varpi$ is the apsidal precession rate, $\Sigma$ is the surface
density of the disk, and $\sigma_r$ is the rms radial velocity, related to the
mean-square eccentricity by $\sigma_r^2=\case{1}{2}r^2n^2\langle
e^2\rangle$. The second number is Toomre's $Q$-parameter (\cite{BT87}) and
$Q>1$ is required for axisymmetric stability. If the precession rate is
determined by the self-gravity of the disk, then generally $S\sim 1$. For
example, a disk with a power-law surface density $\Sigma(r)\propto
r^{-\beta}$ has
\be
S=2{\Gamma(2-\case{1}{2}\beta)\Gamma(\case{1}{2}+\case{1}{2}\beta)\over
\Gamma(\frac{3}{2}-\case{1}{2}\beta)\Gamma(\case{1}{2}\beta)},\quad 0<\beta<3;
\label{eq:spow}
\ee
in this case $S$ is independent of radius and varies only between 1 and 1.0942
in the range $1\le\beta\le 2$.

More generally, if there are other sources of precession then $|S|\gg 1$. In
the rings of Saturn and Uranus, where the precession is dominated by the
planetary quadrupole moment, $S\sim -10^5$. In protoplanetary disks containing
a thin disk of planetesimals embedded in a thicker disk of gas, $S$ will be
roughly the ratio of gas mass to planetesimal mass, $S\sim 10^2$. We shall
find that resonant relaxation is quite different in disks with $S\sim 1$ and
$|S|\gg 1$.

To simplify the discussion, we shall focus on the effects of resonant
relaxation on planets, that is, on bodies much larger than the planetesimals
with which they are interacting. Also, we shall concentrate on resonant
relaxation of the eccentricities, and generally ignore inclination relaxation
except for a brief discussion in \S \ref{sec:disc}. 

The equilibrium distribution of the particle orbital elements in a resonantly
relaxed disk is determined in \S \ref{sec:thermo}. Estimating the rate of
resonant relaxation is a more challenging task.  A variety of tools can be
used for this purpose, depending on the properties of the disk.  We shall
employ secular perturbation theory in \S \ref{sec:lag} to provide analytic
estimates of the rate of resonant relaxation and resonant friction in
disks. We carry out similar calculations using WKB density-wave theory in \S
\ref{sec:wkb}, and numerical normal-mode calculations in \S \ref{sec:num}. The
results are summarized and discussed in \S \ref{sec:disc}.

\section{Thermodynamic equilibrium}

\label{sec:thermo}

\noindent
The equilibrium eccentricity and inclination distribution of the particles in
a resonantly relaxed disk can be determined by the methods of statistical
mechanics. On timescales longer than the resonant relaxation time, the disk
should be in the maximum-entropy state consistent with its total angular
momentum and energy and mass distribution. Thus the equilibrium phase-space
distribution function is (\cite{RT96})
\be
f(\bfr,\bfv,m)=w(E,m)\exp(m\bfb\cdot\bfJ),
\ee
where $E=\case{1}{2} v^2-G\mstar/r$, $\bfJ=\bfr\times\bfv$, and $\bfb$ and
$w(E,m)$ are determined implicitly by the total angular momentum and the
distribution of particles in mass and energy, both of which are
conserved\footnote{In contrast, the equilibrium distribution function for
non-resonant relaxation is
\be
f(\bfr,\bfv,m)=w(m)\exp(-m\beta E+m\bfb\cdot\bfJ),
\ee
where $\beta$ and $\bfb$ are constants.  This cannot be achieved in a
Keplerian potential, since it requires that the density diverges as
$\exp(G\mstar m\beta/r)$ as $r\to0$.}. We orient the coordinate system so that
$\bfb=b\hat {\bf e}_z$ and introduce the actions $(J_c,J,J_z)$, where
$J_c=(G\mstar a)^{1/2}$ is the specific angular momentum of a circular orbit
with semimajor axis $a$, $J=J_c(1-e^2)^{1/2}$ is the specific angular
momentum, $J_z=J\cos I$ is the $z$-component of angular momentum, and $e$ and
$I$ are the eccentricity and inclination.  The conjugate angles are the mean
anomaly, argument of perihelion, and longitude of ascending node
(e.g. \cite{BC61}). Then we have
\be
f(\bfr,\bfv,m)=w[E(J_c),m]\exp(mbJ_z),
\ee
where $E(J_c)=-\case{1}{2}(G\mstar)^2/J_c^2$. The advantage of this form is
that the phase-space volume element is simply $(2\pi)^3dJ_cdJdJ_z$, so the
density of particles in action space is given by
\be
dN=(2\pi)^3w[E(J_c),m]\exp(mbJ_z)dJ_cdJdJ_z.
\ee
If the eccentricities and inclinations are small we may approximate $J$ and
$J_z$ by $J_c(1-\case{1}{2} e^2)$ and $J_c(1-\case{1}{2} e^2-\case{1}{2}
I^2)$, so
\be
dN=W(J_c,m)\exp[-\case{1}{2} mbL(e^2+I^2)]dJ_cde^2dI^2,
\label{eq:ray}
\ee
where $W(J_c,m)=2\pi^3w[E(J_c),m]J_c^2\exp(mbJ_c)$. This is the Rayleigh
distribution (e.g. \cite{LS93}), with the further constraint that the
mean-square eccentricity and inclination at different semimajor axes and
masses are related by $\langle e^2\rangle,\langle I^2\rangle \propto
1/(ma^{1/2})$.

\section{Secular perturbation theory}

\label{sec:lag}

\noindent
We examine the relaxation of a planet of mass $m_P$, semimajor axis $a_P$,
mean motion $n_P$, eccentricity $e_P\ll1$ and longitude of periapsis
$\varpi_P$.  We shall also use the complex eccentricity $z_P\equiv
e_P\exp(i\varpi_P)$. The disk is axisymmetric and composed of $N$ planetesimals
with masses $m_j$ and orbital elements $a_j$, $n_j$, $e_j$, $\varpi_j$, $z_j$,
$j=1,\ldots,N$. The planet is assumed to be much more massive than the
planetesimals, $m_j\ll m_P\ll\mstar$.  Kepler's law states that
$n_j^2a_j^3=n_P^2a_P^3=G\mstar$.
 
We assume that the eccentricities are small in the sense that 
$e_P,e_j\ll |a_j-a_P|/a_P$. Then the secular evolution of the complex
eccentricity is described by Lagrange's equations (\cite{BC61}):
\begin{eqnarray}
\dot z_P&=&ig_Pz_P-i\sum_{j=1}^NA_{Pj}z_j,\nonumber \\
\dot z_k&=&ig(a_k)z_k-iA_{kP}z_P -i\sum_{{j=1}\atop{j\not=k}}^NA_{kj}z_j. 
\label{eq:first}
\end{eqnarray}
Here
\be
A_{kj}={2Gm_j\over n_ka_k^2}P_{kj},\quad 
P_{kj}=P_{jk}={\alpha b^{(2)}_{3/2}(\alpha)\over 8\max(a_k,a_j)},
\quad \alpha={\min(a_k,a_j)\over\max(a_k,a_j)}.
\label{eq:gdef}
\ee
The function $g(a_k)=\dot\varpi_k$ and $g_P=\dot\varpi_P$ are the rates of
apsidal precession of the  disk and the planet; we defer a
discussion of these until \S \ref{sec:prec}. The Laplace coefficient
$b_s^{(m)}(\alpha)$ is defined by
\be
b_s^{(m)}(\alpha)={2\over\pi}\int_0^\pi {\cos m\phi d\phi\over
(1-2\alpha\cos\phi +\alpha^2)^s},
\label{eq:lapl}
\ee 
and we will use the result that
\be
b^{(m)}_{3/2}(\alpha)\simeq {2\over\pi(1-\alpha)^2}\quad\hbox{as }\alpha\to
1.
\label{eq:asymp}
\ee

\subsection{Resonant relaxation}

\noindent
Relaxation arises from the stochastic forces exerted on the planet by the
planetesimal orbits. We concentrate first on disks in which the surface
density is low, in the sense that $|S|\gg1$ (cf. eq. \ref{eq:sdef}). In this
case we can drop the term $-i\sum_{j\not=k}A_{kj}z_j$ which represents mutual
perturbations among planetesimals from the second of equations
(\ref{eq:first}). We also drop the term $-iA_{kP}z_P$ which represents the
forces from the planet on the planetesimals. Thus the planetesimal
eccentricities satisfy $z_k(t)=z_k(0)\exp[ig(a_k)t]$, where $z_k(0)$ is the
initial eccentricity. Substituting this result into the first of equations
(\ref{eq:first}), and assuming that the initial planet eccentricity
$z_P(0)=0$, we obtain
\be
z_P(t)=\sum_jA_{Pj}z_j(0){\exp[ig(a_j)t]-\exp(ig_Pt)\over 
g_P-g(a_j)}.
\ee

We now compute the rate of change of the mean-square planetary eccentricity
$\langle e_P^2\rangle$, where $\langle\cdot\rangle$ denotes an ensemble
average. For axisymmetric disks the longitude of periapsis is uniformly
distributed, so $\langle z_j^\ast(0)z_k(0)\rangle=0$ if $j\not=k$. Using
this result we find 
\be
{d\over dt}\langle e_P^2\rangle=\left\langle z_P^\ast {dz_P\over
dt}\right\rangle+\left\langle z_P{dz_P^\ast\over dt}\right\rangle =2\sum_j
A_{Pj}^2\langle|z_j(0)|^2\rangle{\sin [g_P-g(a_j)]t\over
g_P-g(a_j)}.
\ee
At large times we may use the relation 
$\lim_{t\to\infty}\sin (xt)/x=\pi\delta(x)$, where $\delta(x)$ is the Dirac
delta-function; we also replace the sum over particles by an integral over
semimajor axis. Thus
\be
{d\over dt}\langle e_P^2\rangle={\pi^2\over 4}n_P^2\langle e^2\rangle
{\langle m^2\rangle\over\mstar\langle m\rangle}\int {\Sigma(a)a_Pda\over\mstar}
\alpha^3\left[b_{3/2}^{(2)}(\alpha)\right]^2\delta[g(a)-g_P],
\label{eq:relaxa}
\ee
where $\alpha=\min(a,a_P)/\max(a,a_P)$, $\Sigma(a)$ is the surface density
of the disk, and $\langle m\rangle$ and $\langle m^2\rangle$ are averages over
the mass distribution. In this approximation, relaxation is entirely due to
particles located at the secular resonance $a_s$ where $g(a_s)=g_P$.

\subsection{Resonant friction}

\noindent
To evaluate the resonant friction on the planet, we return to Lagrange's
equations (\ref{eq:first}) and once again drop the term
$-i\sum_{j\not=k}A_{kj}z_j$ representing collective effects within the
protoplanetary disk. Then if we assume that $z_P,z_j\propto \exp(i\omega t)$
we obtain the dispersion relation
\be
g_P-\omega-\sum_j{A_{Pj}A_{jP}\over g(a_j)-\omega}=0.
\label{eq:dispfirst}
\ee
We replace the sum by an integral to obtain
\be
g_P-\omega-{\pi\over 8}n_P{m_P\over\mstar}
\int_L{\Sigma(a)ada\over\mstar}\alpha^3\left[b_{3/2}^{(2)}(\alpha)\right]^2 
{n(a)\over g(a)-\omega}=0.
\label{eq:relaxb}
\ee
The subscript ``L'' is a reminder that the integral for Im$(\omega)>0$ is the
analytic continuation of the integral defined for Im$(\omega)<0$ (the Landau
contour). Equation (\ref{eq:relaxb}) can be solved directly for the
eigenfrequency of the coupled planet-planetesimal system, but it is easier and
more informative to focus on the case where the coupling represented by the
integral in equation (\ref{eq:relaxb}) is weak enough that Re$(\omega)\simeq
g_P$. Then the rate of eccentricity growth or damping from resonant friction
is given by
\be
{de_P^2\over dt}=-2e_P^2\,\hbox{Im}\,(\omega)={\pi\over 4}e_P^2n_P
{m_P\over\mstar}\hbox{Im}
\int_L{\Sigma(a)ada\over\mstar}\alpha^3
\left[b_{3/2}^{(2)}(\alpha)\right]^2 {n(a)\over g(a)-\omega}.
\ee
Since the coupling is weak, Im$(\omega)$ is small. For Im$(\omega)$ small and
negative, we may use the relation
\be
\lim_{\epsilon\to0}{\epsilon\over \epsilon^2+x^2}=
\hbox{sgn}(\epsilon)\pi\delta(x)
\label{eq:delta}
\ee
to replace Im$[(g-\omega)^{-1}]=\hbox{Im\,}\omega/[(g-g_P)^2+(
\hbox{Im\,}\omega)^2]$ by $-\pi\delta(g-g_P)$. Thus we obtain
\be
{de_P^2\over dt}=-{\pi^2\over 4}e_P^2n_P
{m_P\over\mstar}\int{\Sigma(a)ada\over\mstar}n(a)\alpha^3
\left[b_{3/2}^{(2)}(\alpha)\right]^2 
\delta[g(a)-g_P],
\label{eq:relaxc}
\ee
and by Landau's prescription this is the correct expression for small values
of Im$(\omega)$ of either sign. Thus resonant friction always damps the
planet's eccentricity\footnote{This result holds even for spherical systems,
so long as the velocity distribution is predominantly tangential
(\cite{RT96}).}.

Equation (\ref{eq:relaxc}) can be rewritten as
\be
{de_P^2\over dt}=-{\pi^2\over 4}e_P^2n_P
{m_P\over\mstar}{\Sigma(a)a^2\over\mstar}{n(a)\over |dg/d\log a|}
\alpha^3\left[b_{3/2}^{(2)}(\alpha)\right]^2,
\label{eq:relaxd}
\ee
where all quantities are evaluated at the secular resonance $a_s$. 

The planetary eccentricity is in equilibrium when $d\langle e_P^2\rangle/dt$
given by (\ref{eq:relaxa}) plus $de_P^2/dt$ given by (\ref{eq:relaxc}) is
zero. This occurs when
\be
e_P^2={a^{1/2}\over a_P^{1/2}}
{\langle m^2\rangle\over m_P\langle m\rangle}\langle e^2\rangle,
\label{eq:equ}
\ee
where the planetesimal parameters $\langle m\rangle$, $\langle m^2\rangle$,
$a$, $\langle e^2\rangle$ are evaluated at the secular resonance.  Equation
(\ref{eq:equ}) is closely related to the equipartition theorem (\S
\ref{sec:thermo}).

There is an additional complication for a planetesimal disk embedded in a
gaseous disk, as the gas disk may also have a secular resonance with the
planet. The gas resonance will have a different location since the surface
density of the gas disk is larger and its precession rate is modified by
pressure. Equation (\ref{eq:relaxc}) describes the resonant friction from the
gas disk if $\Sigma(a)$ is replaced by the gas surface density and $g(a)$ by
the precession rate of the gas elements.

\subsection{Precession rates}

\label{sec:prec}

\noindent
The location of the secular resonance and hence the resonant relaxation rate
is determined by the rate of apsidal precession of the planet and
the disk particles. Precession can arise from several distinct sources. For
example, the quadrupole moment of the central star induces precession at a
rate
\be
g(a)={3J_2R_\star^2(G\mstar)^{1/2}\over 2a^{7/2}},
\ee
where $R_\star$ and $J_2$ are the radius and dynamical oblateness of the
star. The quadrupole precession is negligible for most protoplanetary disks,
although it dominates the precession in other systems such as planetary
rings.

The planet induces precession in the planetesimals. In Lagrange's 
theory the resulting precession rate of an object with semimajor axis $a_k$
and mean motion $n_k$ is 
\be
g(a_k)={2Gm_P\over n_ka_k^2}N_{kP}, \quad N_{kj}=N_{jk}=
{\alpha b^{(1)}_{3/2}(\alpha)\over 8\max(a_k,a_j)}.
\label{eq:prec}
\ee
Strictly this is the free precession, i.e. the precession that would occur if
the planet orbit were circular. There is also forced precession due to the
planet's eccentricity, which corresponds to the term $-iA_{kP}z_P$ in the
second of equations (\ref{eq:first}).

The planetesimals also induce precession in the planet and each other. By an
obvious extension of equation (\ref{eq:prec}), the free precession of an
object at semimajor axis $a_k$ is given by
\be
g(a_k)=\sum_{j=1}^N{2Gm_j\over n_ka_k^2}N_{kj}.
\label{eq:disc}
\ee
However, the correct interpretation of this result is subtle for a continuous
disk. The subtleties are illustrated by the observation that equation
(\ref{eq:disc}) always predicts prograde free precession ($g(a)>0$), whereas
smooth, continuous, axisymmetric disks usually induce retrograde free
precession (cf. eqs. \ref{eq:sdef} and \ref{eq:spow}).  Moreover in the
continuum limit ($N\to\infty$, $m_j\sim N^{-1}$), the precession rate
predicted by equation (\ref{eq:disc}) diverges as $N^{-1}$, whereas in a
smooth continuous disk the precession rate is finite.

The discrepancy arises because equation (\ref{eq:disc}) assumes that particle
orbits do not cross (more precisely, that $e_j,e_k\ll |a_j-a_k|/a_k$ for all
$j,k$). This assumption is correct for an object with small free
eccentricity that orbits outside the disk (or even within a gap in
the disk), but incorrect if the orbit is embedded in a continuous disk, no
matter how small the free eccentricity. Equation (\ref{eq:disc}) can be
applied to a continuous disk only if the complex eccentricity $z(a)$ is a
smooth function of semimajor axis, so orbits do not cross (i.e. if there are
forced eccentricities but the free eccentricity is zero). 

The simplest fix for this shortcoming is to ``soften'' the gravitational
potential of the particles, by replacing the Laplace coefficient
(\ref{eq:lapl}) by
\be
\wb_s^{(m)}(\alpha)={2\over\pi}\int_0^\pi {\cos m\phi d\phi\over
(1-2\alpha\cos\phi +\alpha^2+\epsilon^2)^s},
\label{eq:lapls}
\ee
where $\epsilon\ll1$. In this case the expressions (\ref{eq:gdef}) and
(\ref{eq:prec}) for $P_{jk}$ and $N_{jk}$ are no longer valid, since they were
derived using recursion relations among unsoftened Laplace coefficients
(\cite{BC61}); they must be replaced by
\begin{eqnarray}
P_{jk} & = & {1\over
4\max(a_k,a_j)}\left[\case{1}{2}\alpha^2{d^2\wb_{1/2}^{(1)}(\alpha)\over
d\alpha^2}+\alpha{d \wb_{1/2}^{(1)}(\alpha)\over
d\alpha}-\wb_{1/2}^{(1)}(\alpha)\right], \nonumber \\
N_{jk} & = & {1\over
4\max(a_k,a_j)}\left[\case{1}{2}\alpha^2{d^2\wb_{1/2}^{(0)}(\alpha)\over
d\alpha^2}+\alpha{d \wb_{1/2}^{(0)}(\alpha)\over
d\alpha}\right].
\end{eqnarray}
The precession rate predicted by equation (\ref{eq:disc}) can then be applied
to a continuous disk as long as the spacing between adjacent particles is
small compared to the softening, $|a_j-a_{j-1}|\ll\epsilon a_j$.

We shall also use $\epsilon$ more generally, to indicate the typical
dimensionless thickness or epicycle size in the disk; for $|1-\alpha|\lesssim
\epsilon$ the Lagrange equations (\ref{eq:first}) are not accurate. 

Yet another possible source of precession is the gravity from the thick gas
disk that generally envelops the planetesimal disk in the early stages of
planet formation.

\subsection{The resonant relaxation rate}

\noindent
To estimate the resonant relaxation rate we first determine the location of
the secular resonance. We write the particle precession rate as
$g(a)=g_0(a)+g_1(a)$, where $g_0$ is the smoothly varying precession rate due
to the disk mass in gas and planetesimals, quadrupole moment of the central
star, etc., while $g_1(a)$ is the precession due to the planet (eq.
\ref{eq:prec}). The precession rate of the planet itself is $g_P=g_0(a_P)$ (unless the disk
properties change sharply at the planet's location, for example if there is a
gap). We then expand $g_0(a)$ in a Taylor series around $a_P$; using equation
(\ref{eq:asymp}) we then have
\be
a_s=a_P-\left({1\over 2\pi}{m_P\over\mstar}{na^2\over
dg_0(a)/da}\right)_{a_P}^{1/3}.
\label{eq:resloc}
\ee
Of course, this result is only valid if $a_P \gg |a_s-a_P|\gg
\epsilon a_P$; in other words the planet mass must be large enough that the
secular resonance is separated by at least the typical epicycle size or disk
thickness ($\sim\epsilon a_P$).  These constraints can be expressed
approximately as
\be
S\epsilon^3\ll {m_P\over m_D} \ll S,
\label{eq:range}
\ee
where $m_D$ is the disk mass.  The rates of resonant relaxation and friction
are then given by equations (\ref{eq:relaxa}) and (\ref{eq:relaxc}):
\begin{eqnarray}
{1\over\langle e_P^2\rangle} {d\langle e_P^2\rangle\over dt} & = &
{(2\pi)^{4/3}\over 3}n{\langle m^2\rangle\over\mstar\langle m\rangle}
{\Sigma a^2\over\mstar}\left|{1\over n}{dg_0\over d\log
a}\right|^{1/3}\left(m_P\over \mstar\right)^{-4/3}, \nonumber \\
{1\over e_P^2} {de_P^2\over dt} & = &
-{(2\pi)^{4/3}\over 3}n{\Sigma a^2\over\mstar}\left|{1\over n}{dg_0\over d\log
a}\right|^{1/3}\left(m_P\over \mstar\right)^{-1/3},
\label{eq:sbig} 
\end{eqnarray}
where all quantities are evaluated at $a_s$.  Note that smaller planets relax
faster, because the resonant radius $a_s$ is closer.

These results have several limitations: (i) We have neglected the
term $-i\sum_{j\not=k}A_{kj}z_j$ in equations (\ref{eq:first}), which
describes collective effects arising from the self-gravity of the 
disk. (ii) The delta functions in equations (\ref{eq:relaxa}) and
(\ref{eq:relaxc}) imply that as $t\to\infty$ all of the relaxation or drag
comes from particles located exactly at the secular resonance. More
generally, at large but finite times the relaxation or damping is due to the
few particles concentrated within an ever-narrowing resonant band $\Delta
a\sim |da/d g_d|/t$; eventually there may be too few particles and too
little mass in the resonant band for the formula to be valid. (iii) A related
concern is that the eccentricities of the particles near resonance will
become so large that nonlinear terms neglected in equations (\ref{eq:first})
will limit the strength of the resonant interactions. 

These concerns can be addressed using the complementary tool of
density-wave theory, which treats the planetesimal disk as a continuous,
collisionless, self-gravitating fluid.

\section{Density-wave theory}

\label{sec:wkb}

\noindent
The disk is taken to have surface density $\Sigma(r)$, radial velocity
dispersion $\sigma_r(r)$, and epicycle frequency $\kappa(r)=n(r)-g(r)$, where
as usual $n(r)$ and $g(r)$ are the mean motion and apsidal precession rate.
We disturb the disk with a surface-density perturbation proportional to
$\hbox{Re}\,\{\exp[i(\int^rk(r)dr+m\phi-\omega t)]\}$, and seek a dispersion
relation that relates the frequency $\omega$ to the radial and azimuthal
wavenumbers $k$ and $m$. In the WKB or tight-winding limit $|kr|\gg 1$ the
dispersion relation for a collisionless thin disk with a Rayleigh distribution
of eccentricities is (e.g. \cite{BT87})
\be
\kappa^2(r)-[mn(r)-\omega]^2-2\pi G\Sigma(r)|k|F\left({\omega-mn(r)\over
\kappa(r)}, {k^2\sigma_r^2(r)\over\kappa^2(r)}\right)=0;
\label{eq:wkb}
\ee
where
\be
F(s,x)={2\over x}(1-s^2)e^{-x}\sum_{j=1}^\infty {I_j(x)\over 1-s^2/j^2},
\ee
and $I_j(x)$ is a modified Bessel function.

For brevity we shall focus on the predictions of density-wave theory for
resonant friction; most of our results apply equally well to resonant
relaxation. Thus we are interested in density waves excited by a planet on an
eccentric orbit. The slowly varying non-axisymmetric component of the planet's
gravitational potential has the form (e.g. \cite{GT80})
\be
\Phi(r,\phi,t)={Gm_Pe_P\over 2\max(a_P,r)}\left(1-\alpha {d\over
d\alpha}\right)b_{1/2}^{(1)}(\alpha)\cos[\phi-\varpi_P(t)],
\label{eq:planpot}
\ee
which corresponds to $m=1$ and frequency $\omega=g_P=\dot\varpi_P$. Since
$|g|,|g_P|\ll n$ in this case, we may simplify equation (\ref{eq:wkb}) by
neglecting terms of order $g^2$:
\be
2n(r)[g_P-g(r)]-2\pi G\Sigma(r)|k|F\left(-1,{k^2\sigma_r^2(r)\over
n^2(r)}\right)= 0,
\label{eq:wkbc}
\ee
or
\be
g_P-g(r)-{2\pi G\Sigma(r)\over\sigma_r(r)}\left[{e^{-x}\over
x^{1/2}}I_1(x)\right]=0,
\label{eq:wkba}
\ee
where $x=k^2\sigma_r^2/n^2$. In the limit $x\ll1$ this simplifies to
\be
g_P-g(r)-{\pi G\Sigma(r)|k|\over n(r)}=0 \quad\hbox{or}\quad 
{g_P\over g(r)}=1-{|kr|\over S(r)};
\label{eq:wkbb}
\ee

In the WKB approximation, the planet excites a density wave satisfying the
dispersion relation (\ref{eq:wkba}).  The wave is excited near the secular
resonance, where $g_P=g(r)$ and $k\simeq 0$. The wave propagates in the region
where $g(r)<g_P$; in the usual case where $g(r)<0$ and $|g(r)|$ decreases
outwards, the wave propagates inwards as a leading spiral ($k<0$); this is the
long-wavelength branch of the dispersion relation. The wave becomes more and
more tightly wound ($|k|$ increases) as it propagates. Eventually the wave
encounters a turning point at $x=0.5841$ (the maximum of the quantity in
square brackets in eq. \ref{eq:wkba}) and thereafter propagates outward,
continuing to become more tightly wound; this is the short-wavelength
branch. The wave-propagation zone may be written
\be
0 < g_P-g(r) < 1.397{G\Sigma(r)\over\sigma_r(r)} = 0.416{n(r)\over Q(r)},
\label{eq:qqq}
\ee
where $Q$ is defined in equation (\ref{eq:sdef}).  As the wave returns to the
secular resonance $|k|\to\infty$ and the wave is damped by wave-particle
resonances.

The gravitational force from the leading spiral wave adds angular
momentum---but almost no energy, since the wave pattern is nearly stationary
in an inertial frame---to the planet orbit. Thus the planetary eccentricity
is damped. The damping rate can be determined from formulae in Goldreich and
Tremaine (1980)\nocite{GT80},
and turns out to be precisely the same as equation (\ref{eq:relaxc}); this is
a special case of the general result that the gravitational torques exerted on
satellites by disks with a given surface-density distribution are largely
independent of the nature of the collective effects in the disk (\cite{GT79}).

This analysis addresses several of the concerns raised at the end of the
previous section. In particular, in the WKB approximation the rate of resonant
relaxation and friction is not affected by the self-gravity of the disk; thus
equations (\ref{eq:relaxa}) and (\ref{eq:relaxc}) for the damping rate remain
valid when collective effects are included. Moreover the collective effects
ensure that a larger part of the disk is gravitationally coupled to the
planet, not just the particles at the secular resonance: for example, equation
(\ref{eq:qqq}) implies that most of the disk inside the secular resonance lies
in the wave zone if $Q\lesssim n/g$; this is in contrast to waves in
non-Keplerian disks or with $m\not=1$, where the much stronger condition
$Q\sim 1$ is needed for a large wave zone.

In fact, the principal limitation of the WKB analysis is that in many disks
the collective effects are so strong that the WKB approximation is invalid.
The ``width'' $\Delta a_s$ of the secular resonance is roughly given by
$\int_{a_s-\Delta a_s}^{a_s} |k|dr=2\pi$, where $k$ is given by the WKB
dispersion relation (\ref{eq:wkbb}).  We find
\be
\Delta a_s=2\pi\left(G\Sigma\over
n|dg/dr|\right)_{a_s}^{1/2}=2\pi^{1/2}a_sS^{1/2}\left({d\log|g|\over d\log
r}\right)^{-1/2}_{a_s}.
\ee
The WKB analysis is only valid if the resonance width is much less than the
distance between the planet and the secular resonance, which in turn requires
\be
{m_P\over m_D} \gg S^{-1/2}.
\label{eq:jjjj}
\ee

The failure of the WKB approximation is even more complete in disks with
$S\sim 1$. Since $\sigma_r\ll nr$ in thin disks, the dispersion relation
(\ref{eq:wkba}) can only be satisfied if the quantity in square brackets is
$\ll1$. This in turn requires that $x\ll 1$ or $x\gg1$; the latter condition
corresponds to the short-wavelength branch of the dispersion relation, which is
rapidly damped by wave-particle interactions (Landau damping). Thus we focus
on the case $x\ll1$, where the dispersion relation is given by
(\ref{eq:wkbb}), which is the same as for a cold disk (eq.
\ref {eq:wkbc} with $\sigma_r\to 0$).  If $g_P/g$ and $S$ are of order unity, 
then (\ref{eq:wkbb}) demands that $|kr|\sim 1$ so the WKB approximation is not
valid, however small the surface density of the disk may be. This situation
contrasts with the case of a non-Keplerian potential or an azimuthal
wavenumber $m\not=1$, where the WKB approximation $|kr|\gg 1$ becomes more and
more accurate as the surface density becomes smaller.  Since the WKB
approximation is invalid, we can only make an order-of-magnitude analytic
estimate of the resonant damping rate. To do this, we set
$\Sigma(a_s)a_s^2\sim m_D$ in equation (\ref{eq:relaxd}), where $m_D$ is the
disk mass. Since the precession is dominated by the disk self-gravity, we set
$|dg/d\log a|\sim n m_D/\mstar$. Setting the Laplace coefficient and
$\alpha_s$ to unity and neglecting other factors of order unity we obtain
\be
{1\over e_P^2}{de_P^2\over dt}\sim-n{m_P\over \mstar}.
\label{eq:relaxe}
\ee
The eccentricity damping rate is independent of the disk mass, except that
when the disk mass becomes smaller than the planet mass the precession is
dominated by the planet so that equation (\ref{eq:relaxe}) is no longer valid.

For a more accurate determination of the rate of eccentricity damping in disks
with $S\sim 1$, we must turn to numerical calculations.

\section{Excitation of normal modes}

\label{sec:num}

\noindent
The preceding discussion suggests that the large-scale, low-frequency, $m=1$
normal modes of thin disks with $S\sim 1$ are largely independent of the
velocity dispersion (eq. \ref{eq:wkbb}). Thus we can determine the response of
the disk to a planet using Lagrange's theory (eq. \ref{eq:first}) as a
discrete approximation to a continuous disk with complex eccentricity $z(a)$.

We first determine the normal modes. If the terms involving the planet are
dropped from the second of equations (\ref{eq:first}), we have
\be
\dot z_k = ig(a_k)z_k-i\sum_{j=1\atop j\not=k}^NA_{kj}z_j, 
\label{eq:firsta}
\ee
where $g(a_k)$ is given by equation (\ref{eq:disc}). This has solution
\be
z_k=\sum_{n=1}^N {b_nc_{k}^n\over\psi_k}\exp(i\omega_nt),
\label{eq:ddd}
\ee
where $\psi_k=(m_kn_k)^{1/2}a_k$, $b_n$ is arbitrary, and $\omega_n$ and
$\bfc^n=(c_1^n,\ldots,c_N^n)$ are the $n^{\rm th}$ eigenvalue and eigenvector
of the matrix $\bfM$ defined by
\be
M_{kj}=g(a_j)\delta_{kj}-{\psi_k\over\psi_j}(1-\delta_{kj})A_{kj};
\ee
that is, $(\bfM-\omega_n{\bf I})\bfc^n=0$. The matrix $\bfM$ is symmetric and
real, so all of the eigenvalues $\omega_n$ are real, and the eigenvectors
$\bfc^n$ can be chosen to be real and orthonormal, that is, $\bfc^n\cdot
\bfc^m=\delta_{mn}$.

Now suppose that a planet is present, with eccentricity
$z_P\propto\exp(i\omega t)$. Since the eigenvectors are complete, the forced 
response of the disk particles may be written in the form
\be
z_k=\sum_{n=1}^N {b_nc_{k}^n\over\psi_k}z_P;
\label{eq:dddd}
\ee
solving the second of equations (\ref{eq:first}) we find
\be
b_n={1\over \omega_n-\omega}\sum_{k=1}^N\psi_kc^n_kA_{kP}.
\label{eq:eeee}
\ee
Using this result to eliminate $z_j$ from the first of equations
(\ref{eq:first}) we obtain the dispersion relation for $\omega$:
\be
g_P-\omega-\sum_{n=1}^N{1\over\omega_n-\omega}\sum_{j,k=1}^N
{\psi_k\over\psi_j}A_{Pj}A_{kP}c_j^nc_k^n=0.
\label{eq:disrel}
\ee
This can be contrasted to equation (\ref{eq:dispfirst}), which was derived by
neglecting the self-gravity of the disk; in this case $c_k^n=\delta_{kn}$,
$\omega_n=g(a_n)$, and equation (\ref{eq:disrel}) reduces to
(\ref{eq:dispfirst}). 

As in the discussion following equation (\ref{eq:dispfirst}) we concentrate on
the case where the planet mass $m_P$ is small compared to the disk mass
$m_D$. We may then proceed as in equations
(\ref{eq:dispfirst})--(\ref{eq:relaxc}) to obtain
\be
{1\over e_P^2}{de_P^2\over
dt}=-2\pi\sum_{n=1}^N\delta(\omega_n-g_P)\sum_{j,k=1}^N
{\psi_k\over\psi_j}A_{Pj}A_{kP}c_j^nc_k^n.
\label{eq:dreal}
\ee
The eccentricity damping arises from resonances between the planet's
precession frequency and the disk eigenfrequencies. In this derivation the
spectral lines of the disk and planet have zero width, but in practice a
number of mechanisms are likely to broaden these lines: (i) evolution of the
planet's mass or semimajor axis during the planet-formation process; (ii)
evolution of the disk mass or surface density due to infall or
angular-momentum transport; (iii) damping of the disk normal modes from
wave-particle resonances or other mechanisms; (iv) the planet's spectral line
is broadened by O(Im$\,\omega$) which is $\sim g_P(m_P/m_D)$ for disks with
$S\sim 1$. Another route to a similar conclusion is to recall that the typical
precession time $g^{-1}$ is likely to be $10^3$--$10^4$ yr (for a disk mass of
$10^{-2}$--$10^{-3}M_\odot$ at a radius of 5 AU), and the likely formation
time for the giant planets is $10^5$--$10^6$ yr, during which the disk
parameters and planetary orbit evolve substantially. Thus we expect the
precessional spectra of the disk and planet to be broadened by at least of
order 1\%. 

To provide a crude representation of this effect---and to help ensure that the
damping rates we obtain are well-behaved---we shall broaden the delta function
into a Lorentz function by replacing $\delta(g_P-\omega_n)$ by
$\pi^{-1}h g_P/[(g_P-\omega_n)^2+h^2g_P^2]$, a replacement which
is exact in the limit $h\to+0$; usually we pick $h=0.1$. The use of a
Lorentzian rather than some other broadening function is arbitrary but
plausible. 

\begin{figure}
\epsscale{0.6}
\plotone{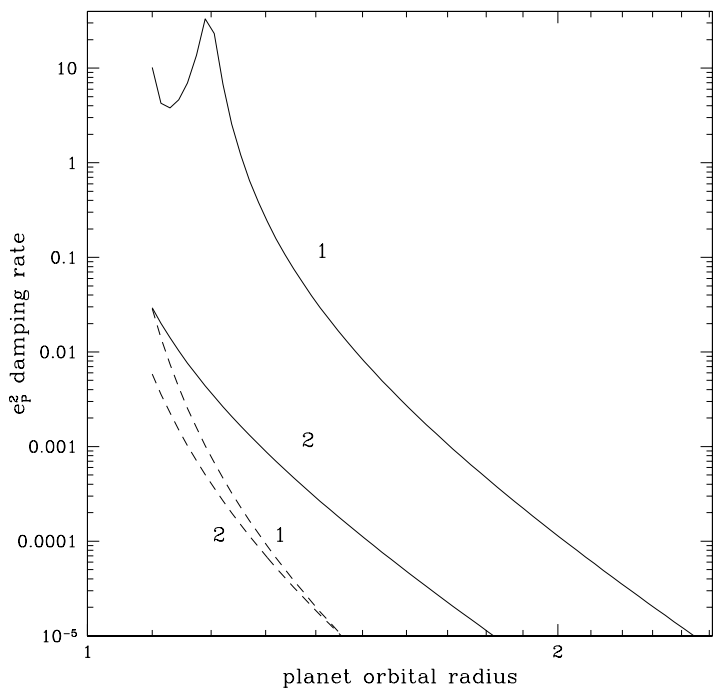}

\caption{The damping rate $d\log(e_P^2)/dt$ for the eccentricity of a planet
orbiting outside a Keplerian disk, as a function of the planet's semimajor
axis. The curves labeled ``1'' are for a sharp-edged disk with surface density
given by equation (\ref{eq:sigone}) and the curves labeled ``2'' are for a
smooth disk (eq. \ref{eq:sigtwo}). The solid curves include the effects of the
disk self-gravity (eq. \ref{eq:dreal}) and the dashed curves do not
(eq. \ref{eq:relaxc}). The units are $G=\mstar=r_0=1$; the damping rate is
proportional to the planet mass $m_P$ and is plotted for $m_P=1$.
\label{fig1}
}

\end{figure}

Figure 1 shows examples of damping rates computed by equation
(\ref{eq:dreal}). We have examined two disk models: the first has sharp edges,
with surface density
\be
\Sigma_1(r)=\left\{ \begin{array}{ll}
                 \Sigma_0(r_0/r)^{1.5}, & r_i<r<r_0 \\
                 0                      & \mbox{otherwise;}
	              \end{array} \right.
\label{eq:sigone} 
\ee
the second is smooth and is obtained by multiplying $\Sigma_1(r)$
by a window function in $\log r$:
\be
\Sigma_2(r)=\Sigma_1(r)\sin^2[\pi\log(r/r_i)/\log(r_0/r_i)].
\label{eq:sigtwo}
\ee
We choose $r_i=0.1r_0$, $h=0.1$, and the results are expressed in units
in which $G=\mstar=r_0=1$. The eccentricity damping rate is proportional to
the planet mass $m_P$ and is plotted for $m_P=1$; in this approximation the
damping rate is independent of the disk surface density $\Sigma_0$.
The planet is assumed to orbit outside $r_0$, and the damping rate is plotted
as a function of the planet's orbital radius. The solid curves are obtained
from equation (\ref{eq:dreal}), which includes the effects of the disk's
self-gravity, while the dashed curves do not (eq. \ref{eq:relaxc}).

The damping rate for the sharp-edged disk exhibits a resonant peak near an
orbital radius of 1.2. The peak arises from a resonance between the planet's
precession rate and the fundamental $m=1$ mode of the
disk (i.e. the normal mode with zero nodes).  At larger orbital radii the
damping rate is determined by the tail of the Lorentz function associated with
this and other normal modes (and thus depends on our arbitrary choice of a
Lorentzian profile with width $h=0.1$). 

The damping rate is lower for the smooth disk; in this case the 
eigenfrequencies of the disk are higher than the precession frequency of the
planet, so the damping rate for a planet outside the disk is everywhere
determined by the tails associated with each resonance. 

These results can be re-stated by writing the heuristic formula
(\ref{eq:relaxe}) as
\be
{1\over e_P^2}{de_P^2\over dt}\sim-fn{m_P\over \mstar};
\label{eq:relaxf}
\ee
then the dimensionless factor $f$ is approximately the same as the ordinate of
Figure 1 for planets orbiting at semimajor axes near the disk edge. The
important feature of Figure 1 is not the numerical damping rates, which depend
on our arbitrary broadening function, but that even modest broadening leads to
substantial damping over a large radius range.

Computing the eccentricity damping rate for a planet embedded in a disk
is more complicated, since (i) the planetary precession rate and hence the
damping will depend sensitively on whether there is an annular gap in the disk
surrounding the planet; (ii) the planetary potential will have higher spatial
frequencies and thus will interact with a richer set of disk normal modes;
(iii) the approximation of small eccentricity is only valid if $e,e_P\ll
|a-a_P|/a_P$, and this is much more difficult to satisfy for nearby
particles; (iv) non-resonant friction due to close encounters between the
planet and disk particles will also play an important role in eccentricity
evolution (see \S \ref{sec:disc} below).

\section{Discussion}

\label{sec:disc}

\noindent
We have investigated resonant gravitational relaxation, which arises in
particle disks orbiting in a nearly Keplerian potential. Resonant relaxation
does not affect semimajor axes, and tends to produce a Rayleigh distribution
in eccentricity and inclination, with mean-square eccentricity or inclination
inversely proportional to mass (eq. \ref{eq:ray}).

The nature of resonant relaxation depends strongly on whether apsidal
precession is dominated by the self-gravity of the disk, a condition which is
parametrized by $S(r)$ (eq. \ref{eq:sdef}). If other mechanisms
dominate the precession ($|S|\gg 1$) then resonant relaxation requires that a
secular resonance is present in the disk, which in turn requires
$m_P/m_D\lesssim S$, where $m_P$ and $m_D$ are the planet and disk masses (eq.
\ref{eq:range}). If the resonance is present, the rate of damping of the
planet's eccentricity from resonant friction is roughly (eq. \ref{eq:sbig})
\be
{1\over e_P}{de_P\over dt}\sim -n\left(m_D\over
\mstar\right)^{4/3}S^{1/3}\left(m_P\over\mstar\right)^{-1/3}, 
\ee
and the rate of stochastic excitation of the eccentricity is
\be
{1\over\langle e_P^2\rangle} {d\langle e_P^2\rangle\over dt}=
{\langle m^2\rangle\over\mstar\langle m\rangle}\left(\mstar\over
m_P\right)\left|{1\over e_P}{de_P\over dt}\right|. 
\ee
These formulae also require that (i) the planet mass is large enough that the
secular resonance is separated from the planet by more than the characteristic
disk thickness or epicycle size (eq. \ref{eq:range}); (ii) the separation
exceeds the characteristic width of the resonance caused by the disk
self-gravity, $m_P/m_D \gtrsim S^{-1/2}$ (eq. \ref{eq:jjjj}).

If on the other hand the apsidal precession is determined by the disk mass
($S\sim 1$), the planet interacts with the large-scale low-frequency normal
modes of the disk rather than with particles near the secular
resonance. Calculating the rate of resonant relaxation in this case is harder
(cf. \S \ref{sec:num}), especially if the planet is embedded
in the disk.  For heuristic purposes the rate of eccentricity decay from
resonant friction may be written as (eq. \ref{eq:relaxe})
\be
{1\over e_P^2}{de_P^2\over dt} \sim -n{m_P\over \mstar}.
\label{eq:kkkk}
\ee
A key point is that the damping rate is independent of the disk mass $m_D$ so
long as the disk mass determines the precession rate. 

A major uncertainty in equation (\ref{eq:kkkk}) is whether the planetary
precession rate is in resonance with one of the disk eigenfrequencies; this
usually requires either a continuous spectrum of eigenfrequencies or that
dissipative or evolutionary processes broaden the discrete disk
eigenfrequencies so that they overlap. Because the characteristic frequency of
the disk modes is of order $n(m_D/\mstar)\ll n$, even slow evolutionary or
damping processes cause considerable broadening. An important next step
will be to understand the low-frequency, large-scale $m=1$ eigenmodes of
nearly Keplerian disks.

For comparison, the non-resonant eccentricity damping rate for a planet
embedded in a disk may be written as (\cite{SW88}, \cite{LS93})
\be
{1\over e_P^2}{de_P^2\over dt} \sim -n{m_P\over \mstar}{m_D\over
\mstar}\left(v_c\over v_r\right)^4\ln\Lambda,
\ee
where $v_r$ is the rms random velocity of the planetesimals, $v_c$ is the
circular speed, and $\ln\Lambda$ is the Coulomb logarithm. As described in \S
\ref{sec:intro}, this result is valid so long as $v_r\gg nr_H$ where $r_H$ is
the planet's Hill radius (\cite{I90}). 

If a planet orbits at a distance $\Delta a$ from a narrow ring, the
non-resonant damping rate becomes (\cite{GT80}, eq.  31)
\be
{1\over e_P^2}{de_P^2\over dt} \sim -n{m_P\over \mstar}{m_D\over
\mstar}\left(a_p\over |\Delta a|\right)^5.
\ee

These crude formulae suggest that either resonant or non-resonant eccentricity
damping can dominate in a given disk, depending on the disk mass, velocity
dispersion, and other parameters.  In typical protoplanetary disks,
non-resonant damping probably dominates if the planet is embedded in the disk.
However if there is a gap $\Delta a/a_P\gtrsim (m_D/\mstar)^{1/5}$ then
resonant damping may be stronger.

Resonant inclination relaxation is similar to resonant eccentricity
relaxation. The principal difference is that the precession due to the
self-gravity of the disk is $g \sim n m_D/\mstar (v_c/v_r)$, stronger by a
factor of order $v_c/v_r$.  Thus the minimum value of the parameter $|S|\sim
v_c/v_r\gg 1$. 

Resonant relaxation is a complex process whose efficiency depends sensitively
on properties such as the precession rate, the presence of gaps around
planets, etc. (although it is much less sensitive to the rms random velocity
in the disk than non-resonant relaxation). Moreover many of the tools that are
usually employed to investigate disk relaxation (two-body scattering,
Fokker-Planck approximation, Hill's problem, etc.) are too crude to describe
resonant relaxation. Nevertheless this process is likely to play an important
role in the evolution of a variety of nearly Keplerian disks, including both
protoplanetary disks and other objects such as stellar disks surrounding black
holes at the centers of galaxies.

\acknowledgements

I thank Kevin Rauch for comments and suggestions. This research was supported
in part by NASA Grant NAG5-7310.

\end{document}